\documentclass[12pt]{article}
\usepackage{amsmath}
\usepackage{amsfonts}
\usepackage{amssymb}
\usepackage{subfig}
\usepackage[countmax]{subfloat}
\usepackage{graphicx}
\usepackage{setspace}
\usepackage{soul}
\usepackage{xcolor}
\usepackage[numbers,sort&compress]{natbib}
\RequirePackage[colorlinks=black,citecolor=blue,urlcolor=blue]{hyperref}

\numberwithin{equation}{section}
\newcommand{\bzet}{{\boldsymbol \zeta}}
\def\bbR{{\mathbb R}}

\def\bbH{{\mathbb H}}

\def\bz{{\boldsymbol z}}

\def\bk{{\boldsymbol k}}

\def\cH{{\mathcal H}}
\def\cK{{\mathcal K}}

\graphicspath{%
    {converted_graphics/}
    {/}
}
\begin{document}

\title{Classical and quantum harmonic oscillators\\subject to a time dependent force} 
\author{Henryk Gzyl,\\
\noindent 
Centro de Finanzas, IESA, Caracas.\\
henryk.gzyl@iesa.edu}

\date{}
 \maketitle

\setlength{\textwidth}{4in}
\vskip 1 truecm
\baselineskip=1.5 \baselineskip \setlength{\textwidth}{6in}
\begin{abstract}
In this work we address the problem of the quantization of a simple harmonic oscillator that is perturbed by a time dependent force. The approach consists of removing the perturbation by a canonical change of coordinates. Since the quantization procedure uses the classical Hamiltonian formalism as staring point, the change of variables is carried out using canonical transformations, and to transform between the quantized systems the canonical transformation is implemented as a unitary transformation mapping the states of the perturbed and unperturbed system onto each other.  
 \end{abstract}



\section{Introduction and Preliminaries}
When confronted with a classical or quantum system subject to a perturbation, one possible approach to describe it consists in removing the perturbation by means of a change of coordinates. So the first problem to consider is to determine the change of coordinates that simplifies the description of the system. Here, the correct way is dictated by the fact that the time evolution of a quantum system is described by a Hamiltonian operator which is obtained from the classical (non quantum) Hamiltonian by an application of the Bohr-Sommerfeld rules. Thus, if we want to remove the effect of a perturbation by means of a change of variables, it is natural to do it first in the classical version of the system and then transport the procedure to the quantized version. It so happens that the canonical transformation that maps the perturbed oscillator onto the simple oscillator can be implemented as a unitary transformation between the corresponding quantized versions of the two systems. It is the aim of this paper to work this out explicitly and thus obtain the time evolution of the perturbed system in terms of that of the unperturbed one.

The  study of time dependent Hamiltonian systems in general and harmonic oscillators with time dependent Hamiltonians is not new. Much a attention has been devoted to systems with time dependent quadratic Hamiltonians. To cite but a few references, consider \cite{U} in which the quantum perturbed oscillator appears as part of s system to detect gravitational waves. In \cite{Li} the Schr\"{o}dinger equation is solved directly. Time dependent quadratic Hamiltonians, which include the cases of the damped harmonic oscillator and/or the harmonic oscillator with a time dependent frequency, have been treated by a variety of techniques. For example, \cite{Ch} uses the Hamilton-Jacobi equation to obtain a solution to the Schr\"{o}dinger equation. The solution to the Hamilton-Jacobi equation is the generating function of a canonical transformation that brings the system to its initial state.
Another approach is based on the construction of invariants of motion (the Ermakov-Lewis invariant) using a  variety of techniques, including canonical transformations. See for example \cite{L},  \cite{SCh}, \cite{KO}, and \cite{URF} (in which  references to applications are given).  Other approaches involve more advanced group theoretic or algebraic approaches using symplectic geometry: \cite{S}, \cite{St}, \cite{SR}, \cite{GGV},  \cite{BNASH} for a sample of works using these approaches. Anyway, all of these differ from the approach developed here, which consists of using a the machinery of canonical transformations and their unitary representations to solve the quantum mechanical problem.

To establish notations, in the remainder of this section we solve the equations of motion of the forced harmonic oscillator within the Hamiltonian formalism, and then see how the solution can be described as the superposition of the solution without the forcing term plus describing the motion of a  system under the action of the ``external'' forcing. 

In Section 2 we explain how to relate the two descriptions by means of a canonical transformation. This the first step to relate the quantized versions on the forced and the non forced oscillators. The second is to implement the canonical transformation as a unitary transformation between the corresponding Hilbert spaces. We do that in Section 3, where we prove that the unitary transformation maps the position, momentum and Hamiltonian operators as in the classical case. On the one hand this means that the quantization rules behave consistently under canonical transformations, and on the other, that we can transform  the solutions of the corresponding Schr\"odinger equations onto each other. There we also indicate how we can compute transition probabilities using the canonical mapping.

\subsection{The Hamiltonian description of the forced harmonic oscillator}
The harmonic oscillator subject to a time dependent force independent of its position is a simple mechanical system. The Hamiltonian from which the dynamics of the forced oscillator is obtained is $H(x,p)=p^2/2m+m\omega^2x^2/2-xk(t).$ When $\omega=0$ the Hamiltonian describes a particle in spatially constant but time dependent force, while when $k(t)=k$ is time independent, the Hamiltonian describes a particle under the action of a harmonic force plus a constant force. 

In the Hamiltonian formalism the equations of motion are obtained from the Hamiltonian $H(x,p)$ as follows:
\begin{equation}\label{H1}
\frac{d}{dt}{x \atopwithdelims ( ) p} = {\frac{\partial H}{\partial p} \atopwithdelims ( ) -\frac{\partial H}{\partial x}} = {p \atopwithdelims ( ) -\omega^2 x+k(t)} = 
\left(\begin{array}{cc}
	0 & 1/m\\
        -m\omega^2 & 0\end{array}\right){x \atopwithdelims ( ) p} +{0\atopwithdelims ( )k(t)} =\bbH_0{x \atopwithdelims ( ) p} +{0\atopwithdelims ( )k(t)}.	
\end{equation}
The initial conditions are $x(0)=x_0,p(0)=p_0.$ This is a common textbook example. It appears as a mathematical model of many physical systems: from electrical circuits to charged particles taken  out of equilibrium by the action an electric fields. See \cite{FLS}, \cite{G} or \cite{A} for example. The solution to that system is simple to obtain and it is given by:

\begin{equation}\label{H2}
{x(t) \atopwithdelims ( ) p(t)} = U(t){x_0\atopwithdelims ( )p_0} + \int_0^tU(t-s){0\atopwithdelims ( )k(s)}ds.		
\end{equation}
Here it is clear that the time dependence of the forcing term is arbitrary, subject to a mathematical requirement, namely, that the integrals on the right hand side of (\ref{H2}) are defined. See the concluding remarks section for more on this issue. In (\ref{H2}) we introduced the following notations:
\begin{equation}\label{H2.1}
U(t) = \left(\begin{array}{cc}
	cos(\omega t) & \frac{1}{\sqrt{m}\omega}sin(\omega t)\\
        -\omega\sqrt{m} sin(\omega t) & cos(\omega t)\end{array}\right),
\end{equation}

\begin{figure}[h]
\centering
\subfloat[$\omega=2\pi/100$]{\includegraphics[width=2.00in,height=2.00in]{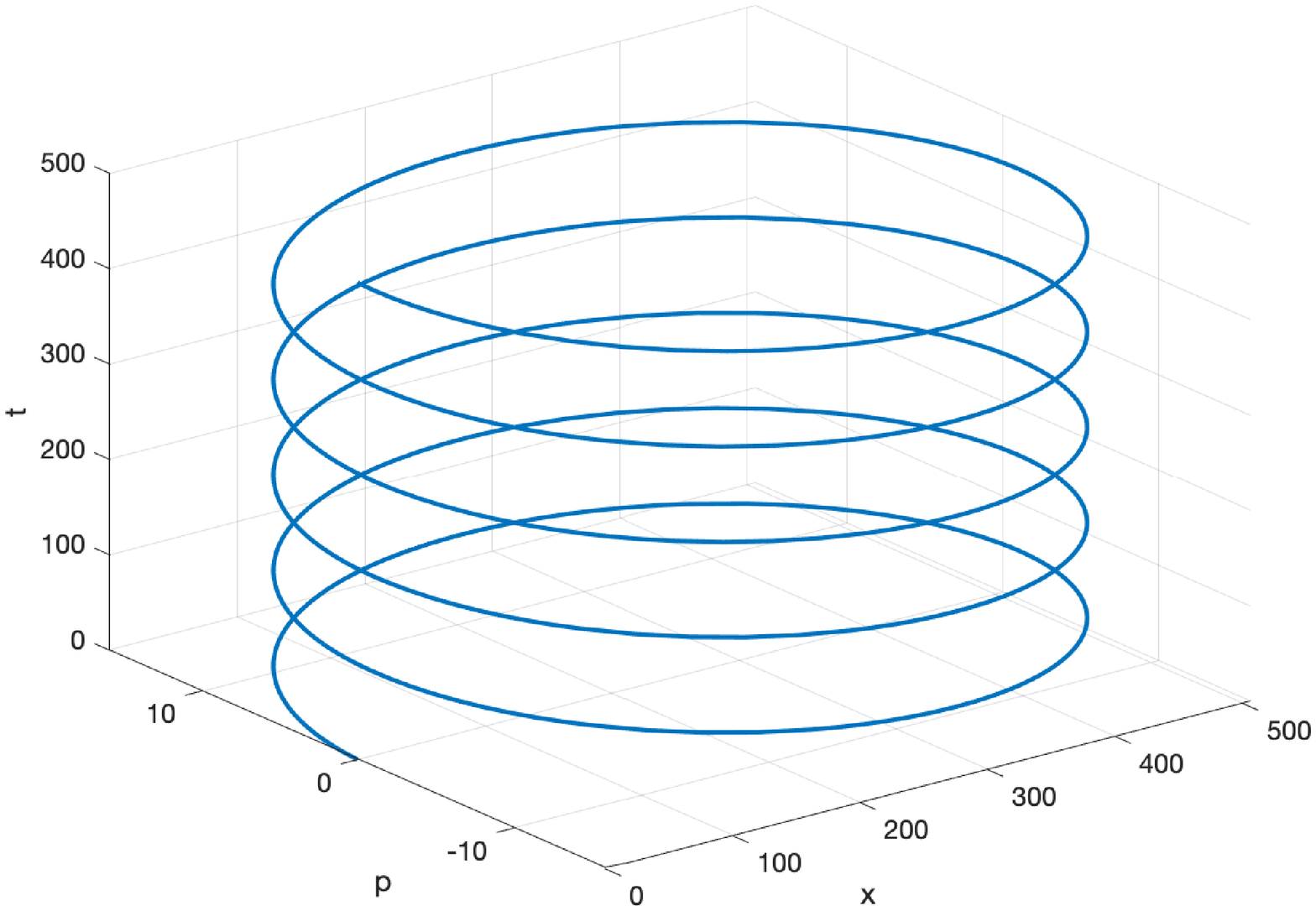}\label{fig1}} 
\subfloat[$\omega=2\pi$]{\includegraphics[width=2.00in,height=2.00in]{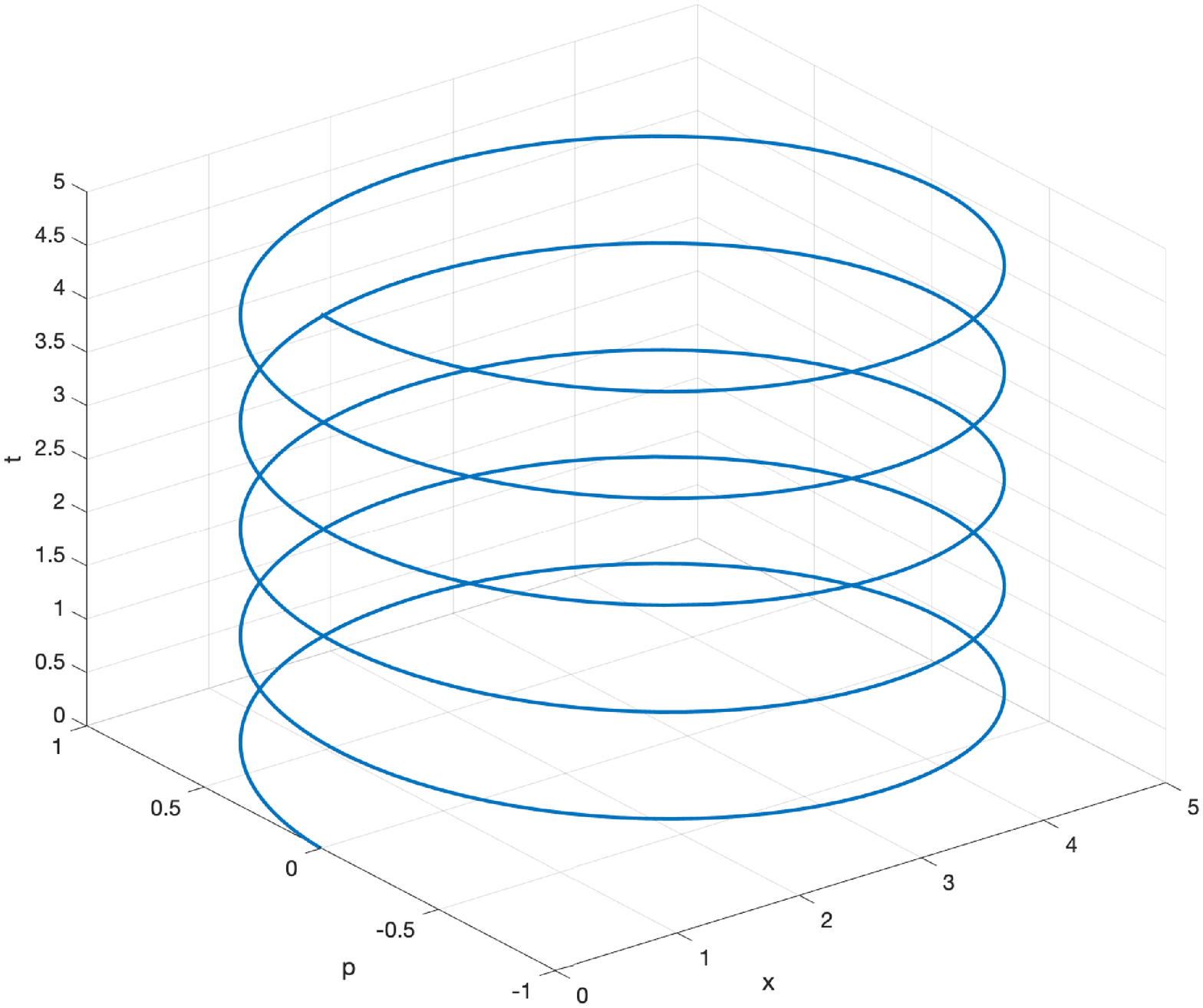}\label{fig2}}
\subfloat[$\omega=200\pi$]{\includegraphics[width=2.00in,height=2.00in]{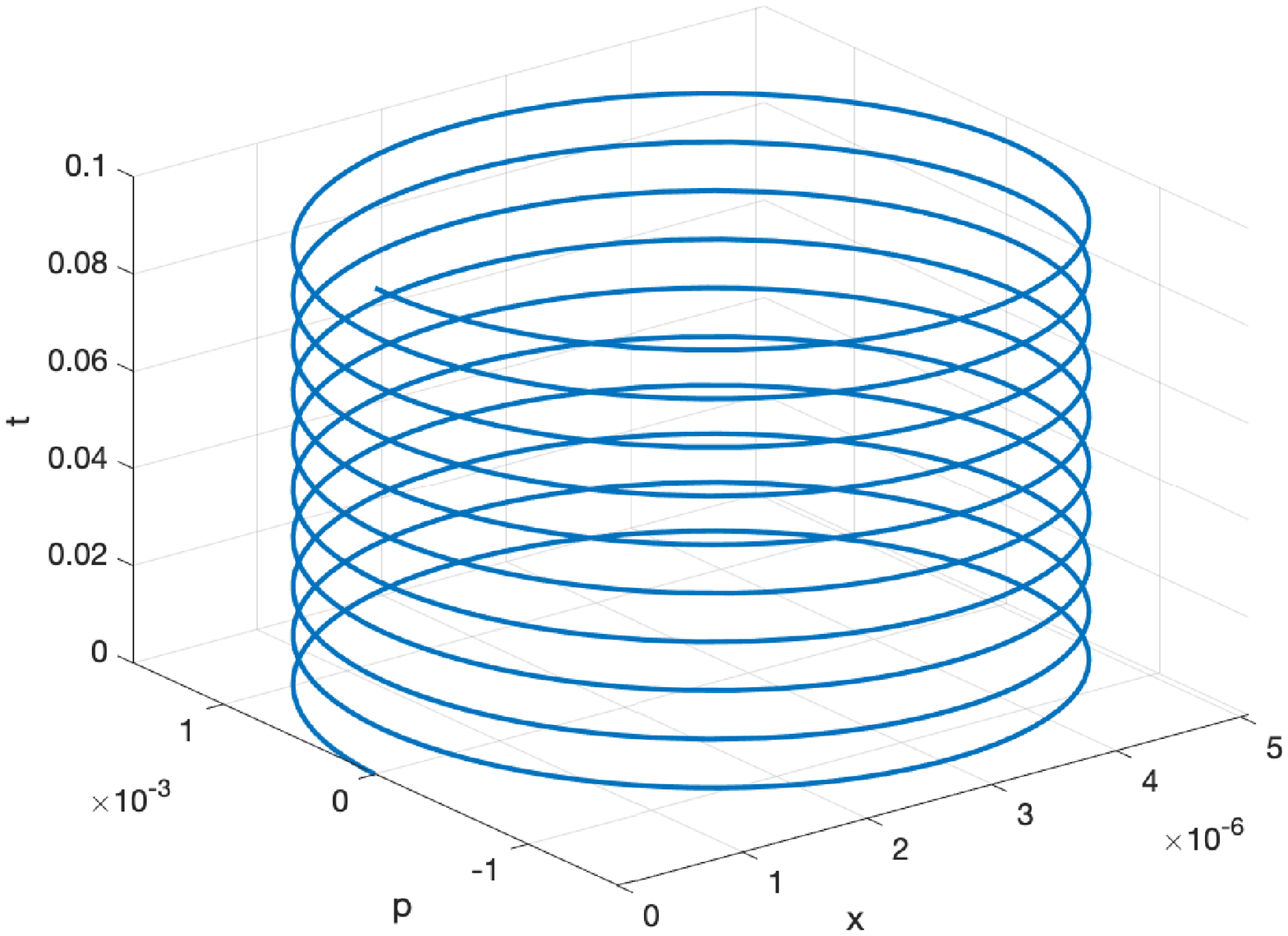}\label{fig3}}
\caption{Motion of the laboratory system}
\label{figA}
\end{figure}

Note as well that $U(t+s)=U(t)U(s)$ or $U(t-s)=U(t)U(-s)$ for all $s,t.$ 
To simplify the notations in what comes below, let us denote the coordinates $(x,v)$ by $\bz$ (thought of as column vector), and denote by $\bk(s)$ the transpose of $(0,k(s)).$ With those notations, write the solution (\ref{H2}) to the system (\ref{H1}) as
\begin{equation}\label{H3}
\bz(t) = U(t)\bz(0) + \int_0^tU(t-s)\bk(s)ds = \bz_h(t) + \bz_{nh}(t).
\end{equation}
The subscript $h$ stands for {\it homogeneous} and $nh$ stands for {\it non-homogeneous}. 
If we put $H_0(x,p)=\frac{1}{2}\big(m\omega^2x^2+p^2/m\big),$ then $\bz_h(t)$ solves (\ref{H1}) with $H_0(x,p)$ instead of $H(x,p).$ Or, if you prefer, $\bz_{nh}(t)$ is just the particular solution to (\ref{H1}) with zero initial conditions. It is also easy to see that the matrix $\bbH_0$ introduced in (\ref{H1}) satisfies
\begin{equation}\label{inv1}
U^\dag(t)\bbH_0 U(t) = \bbH_0.
\end{equation}
Using this we have the following geometric way of visualizing the solutions is as ellipses with center moving according to $\bz_{nh}(t)$. This follows from the fact that
$$\langle\big(\bz(t)-\bz_{nh}(t)\big),\bbH_0\big(\bz(t)-\bz_{nh}(t)\big)\rangle=\mbox{constant}=\langle\bz(0),\bbH_0\bz(0)\rangle.$$
That is, we might think of the two terms in the right hand side of (\ref{H3}) as follows: Interpret $\bz_{nh}(t)$ as the motion of the center of coordinates of a ``laboratory'' that undergoes a non-uniform motion, and interpret $\bz_h(t)$ as the motion with respect to a system of coordinates in which there is no external force, that is of the motion described in a system of coordinates moving with the laboratory. To visualize the motion of the laboratory system, consider the three ellipses displayed in the panels of Figure \ref{figA}. These correspond to $\bz_{nh}(t)$ when $k(t)=K$ is constant in time. In this case we have $(\frac{K}{\omega})\big(\frac{1-cos(\omega t)}{\omega}, sen(\omega t)\big),$ which is an ellipse. Notice that it stretches out when $\omega\to 0$ and shrinks when $\omega\to\infty.$ 
Notice that the scales in the time axis and in the coordinates axes are very different. Had we plotted them on the same scale, the first ellipse (\ref{fig1}) would be very elongated compared to (\ref{fig2}), while the third (\ref{fig3}) would be compressed almost on top of the origin of coordinates.

\section{The canonical transformations}
To perform the change of frame of reference mentioned at the end of the previous section in such a way that the Hamiltonian equations of motion are preserved, we have to consider canonical transformations. We shall think of the coordinates $\bz=(x,p)^t$ as coordinates in the laboratory system and the new coordinates $\bzet=(\xi,\eta)$ as the position and momentum in the moving coordinate system. Let us denote the coordinates of $\bz_{nh}(t)$ by $x_{nh}(t)$ and $p_{nh}(t)=m\dot{x}_{nh}(t),$ where as customary, the dot stands for the derivative with respect to time. 

Let us write $H(x,p)=(p^2/m+m\omega^2x^2)/2-xk(t)$ and $K(\xi,\eta)=(\eta^2/m+m\omega^2\xi^2)/2$ for the two Hamiltonian functions of interest. The passage from one to the other by means of a canonical transform can be achieved in several ways, but we shall consider the following cases  when implementing them as unitary transformations. We put
\begin{eqnarray}\nonumber\\
\mbox{Either}\;\;\;F_1(x,\eta,t) = \big(x-x_{nh}(t)\big)\big(\eta+\dot{x}_{nh}(t)\big)+G(t)\label{CT1}\\
\mbox{or}\;\;\;\;F_2(\xi,p,t) = \big(\xi+x_{nh}(t)\big)\big(p-\dot{x}_{nh}(t)\big)+G(t)\label{CT2}
\end{eqnarray}
depending on which coordinates we want to regard old or new. Which one we consider will depend on which solution of the two possible Schr\"odinger equations we want to transform onto which. The equations that relate the old and new coordinates are (see \cite{A} or \cite{G}):
\begin{eqnarray}\nonumber\\
\xi(t) = \frac{\partial F_1}{\partial\eta}\label{CT3a}\\
p(t) = \frac{\partial F_1}{\partial x}\label{CT3b}\\
K = H + \frac{\partial F_1}{\partial t}\label{CT3c}.
\end{eqnarray}
Note that the transformation equations yield that $\xi(t)=x(t)-x_{nh}(t)$ and $p=\eta+\dot{x}_{nh}(t).$
It is understood that in (\ref{CT3c}) the partial derivation with respect to $t$ is carried out and the old coordinates are substituted for the new after solving (\ref{CT3a})-(\ref{CT3b}). To see that $H(x,p)$ becomes $K(\xi,\eta)$ when we apply (\ref{CT3a})-(\ref{CT3b})-(\ref{CT3c}) to (\ref{CT1}) we have to make use of the fact that
\begin{equation}\label{CT4a}
\ddot{x}_{nh}(t) + m\omega^2x_{nh}(t)-k(t)=0
\end{equation}
and that $G(t)$ is to be chosen so that
\begin{equation}\label{CT4b}
\dot{G}(t) -\dot{x}_{nh}(t)^2/2m + m\omega^2x_{nh}(t)^2/2 - x_{nh}(t)k(t) = 0.
\end{equation}
A similar procedure is followed for the passage from $K(\xi,\eta)$ to $H(x,p).$

\section{Representation of the canonical transformations and quantization}
Our starting point consists of two classical systems, with  phase spaces labeled by $(x,p)$ and $(\xi,\eta),$ and with dynamics determined by the Hamiltonians $H(x,p)$ and $K(\xi,\eta).$ We shall denote by $\Psi(x)$ and $\Phi(\xi)$ the states in the Schr\"dinger representation, and we shall suppose that they are at least twice continuously differentiable, square integrable, complex valued functions such that all the integration by parts necessary to verify the identities presented below are valid. Let us denote by $\cH$ and $\cK$ the corresponding state spaces provided with the usual scalar product. 

As the canonical transformations involve the momentum variables of one system and coordinate variables of the other, the unitary transformations are defined to act on the momentum representation of the states to yield states in the coordinate representations. The description of the states in terms of the momentum representation is given by taking the Fourier transforms, that is: 
\begin{equation}\label{MR}
\Psi(p) = \frac{1}{\sqrt{2\pi}}\int e^{-ipx}\Psi(x)dx\;\;\;\;\mbox{and}\;\;\;\;\Phi(\eta) = \frac{1}{\sqrt{2\pi}}\int e^{-i\eta\xi}\Phi(\xi)d\xi.
\end{equation}
Consider (\ref{CT1})-(\ref{CT2}). In each case the transform depends on the ``new'' momentum and the ``old'' coordinates. Thus the unitary transformations induced by $F_1$ and $F_2$ go in the opposite directions and are defined as follows.
\begin{eqnarray}\nonumber\\
\tilde{\Phi}(x,t) = U_{F_1}(t)\Phi(x,t) = \frac{1}{\sqrt{2\pi}}\int e^{iF_1(x,\eta)}\Phi(\eta,t)d\eta,\label{UT1a}\\
\tilde{\Psi}(\xi,t) = U_{F_2}(t)\Psi(x,t) = \frac{1}{\sqrt{2\pi}}\int e^{iF_2(\xi,p)}\Psi(p,t)dp.\label{UT1b}
\end{eqnarray}
The tilde is a notational reminder of the fact that the state is obtained by applying the canonical transform. Keep in mind that the transformation is time dependent and applied to the state dynamically. Note as well that at $t=0$ the transforms reduce to the connection between the momentum and the coordinate representations. The explicit computation of these transforms is simple:
\begin{eqnarray}\nonumber\\
\widetilde{\Phi}(x,t) = e^{i\theta(x,t)}\Phi(x-x_{nh}(t),t),\label{UT2a}\\
\widetilde{\Psi}(\xi,t) = e^{i\theta'(\xi,t)}\Psi(\xi+x_{nh}(t),t).\label{UT2b}
\end{eqnarray}
Here we put
\begin{eqnarray}\nonumber\\
\theta(x,t) = \big(x-x_{nh}(t)\big)\dot{x}_{nh}(t) + G(t)\label{CTA}\\
\theta'(\xi,t) = -\big(\xi+x_{nh}(t)\big)\dot{x}_{nh}(t) + G(t).\label{CTB}
\end{eqnarray}
Clearly the transformations are unitary. Let us now verify that the position and momentum operators are transformed consistently with the correspondence rules. 

\subsection{Transformation of the position and momentum operators}
Let us begin by recalling that using (\ref{MR}) one verifies that 
$$\hat{\xi}\Phi(\eta) = -i\frac{\partial}{\partial \eta}\Phi(\eta)\;\;\;\mbox{and}\;\;\;\hat{\eta}\Phi(\eta) = \eta\Phi(\eta).$$
Now, applying $U_{F_1}$ to $\Phi$ as in (\ref{UT1a}) and (\ref{UT1b}) we obtain that the position operator transforms as:
\begin{equation}\label{TE1}
U_{F_1}\big(\hat{\xi}\Phi\big)(x,t) = \big(x - x_{nh}(t)\big)\widetilde{\Phi}(x,t) = \big(\hat{x}-x_{nh}(t)\big)\tilde{\Phi}(x,t),
\end{equation}
whereas the momentum operator transforms as:
\begin{equation}\label{TE2}
U_{F_1}\big(\hat{\eta}\Phi\big)(x,t) = \big(-i\frac{\partial}{\partial x} - \dot{x}_{nh}(t)\big)\widetilde{\Phi}(x,t) = \big(\hat{p}-\dot{x}_{nh}(t)\big)\widetilde{\Phi}(x,t).
\end{equation}
To conclude, keep in mind that the Hamiltonian operators in the ``laboratory'' system and  in the ``accelerated'' system are given by
\begin{equation}\label{HO}
\cH = -\frac{1}{2m}\frac{\partial^2}{\partial x^2} + \frac{m\omega^2}{2}x^2-xk(t)\;\;\;\mbox{resp.}\;\;\;\;\cK = -\frac{1}{2m}\frac{\partial^2}{\partial \xi^2} + \frac{m\omega^2}{2}\xi^2.
\end{equation}

\subsection{The evolution in time transforms consistently}
Let us now verity that the time evolution in with respect to one Hamiltonian transforms into time evolution with respect to the other Hamiltonian. The result drops out of the fact that
\begin{equation}\label{SE1}
i\frac{d}{dt}\tilde{\Psi}(\xi,t) = i\frac{d}{dt}\big(U_{F_2}\Psi\big)(\xi,t) = \big(i\frac{d}{dt}U_{F_2}\big)\Psi(\xi.t) + U_{F_2}\big(i\frac{d}{dt}\Psi\big)(\xi,t).
\end{equation}
Actually, a look at the definition in (\ref{UT1a}) and its effect on $\Psi$ explicitly shown in (\ref{UT2b}), shows that the claim will follow from the fact that
\begin{eqnarray}\label{SE2}\nonumber
i\frac{d}{dt}\tilde{\Psi}(\xi,t) = i\frac{d}{dt}\big(e^{i\theta'(\xi,t)}\Psi(\xi+x_{nh}(t),t)\big)\nonumber\;\;\;\;\;\;\;\;\;\\\
 = i\frac{d}{dt}\big(e^{i\theta'(\xi,t)}\big)\Psi(\xi+x_{nh}(t),t) + e^{i\theta'(\xi,t)}\big(i\frac{d}{dt}\Psi(\xi+x_{nh}(t),t)\big).
\end{eqnarray}
Notice that 
$$\big(i\frac{d}{dt}\Psi(\xi+x_{nh}(t),t)\big) = i\dot{x}_{nh}(t)\frac{\partial}{\partial x}\Psi(\xi+x_{nh}(t),t) +
\cH\Psi(\xi+x_{nh}(t),t).$$
It is just a matter of substituting (\ref{TE1})-(\ref{TE2}), making use of (\ref{CT3a})-(\ref{CT3c}), and the fact that neither $x_{nh}(t)$ nor $\dot{x}_{nh}(t)$ involves $x$ or $p,$ to verify that all necessary cancellations take place to conclude that when $\Psi(x,t)$ satisfies Schr\"odinger's equation in the laboratory system, then
\begin{equation}\label{SE3}
i\frac{d}{dt}\tilde{\Psi}(\xi,t) = \cK\tilde{\Psi}(\xi,t),\;\;\;\mbox{with}\;\;\;\tilde{\Psi}(\xi,0)=\Psi(\xi,0)=\Psi(x,0).
\end{equation}
That is, $\tilde{\Psi}(\xi,t)$ satisfies Schr\"odinger's equation in the moving system. An exactly analogous argument proves that if $\Phi(\xi,t)$ satisfies Schr\"odinger's equation with Hamiltonian operator $\cK,$ then $\tilde{\Phi}(x,t)=\big(U_{F_2}\Phi\big)(x,t)$ satisfies "Schr\"odinger's equation with Hamiltonian $\cH.$ Next, we use these results to compute the transition probabilities between the eigenstates of the oscillator induced by the time dependent perturbation.

\subsection{Computation of transition probabilities}
Suppose that at $t=0$ the oscillator was in the $n$th eigenstate $\Psi_n$of energy $E_n$ and the perturbation is turned on. We are interested in computing the probability of finding the oscillator in some other eigenstate $\Psi_m$ as time passes by. That is we want to compute
$$P_{n,m}(t) = |\langle\Psi_m,\Psi_n(t)\rangle|^2.$$
From the closing comment in the previous section we know that the state at current time $t$ that evolved from the eigenstate $\Psi_n(x)=\Phi_n(\xi)$ at time $0$ is, according to (\ref{UT2a}), given by $\widetilde{\Phi_n}(x,t) = e^{-i\theta(x,t)}\Phi_n(x-x_{nh}(t),t)$ and that $\Phi_n(x-x_{nh}(t),t)=\exp(-iE_nt)\Phi_n(x-x_{nh}(t)).$ From (\ref{CTA}) we obtain
$$\langle\Psi_m,\Psi_n(t)\rangle = e^{-iE_nt}e^{-iG(t)}\int\Psi_m(x)e^{-i(x-x_{nh}(t))\dot{x}_{nh}(t)}\Psi_n(x-x_{nh}(t))dx.$$
The phase in front of the integral disappears when we consider the absolute value, so we drop it. A change of variables in the integral yields
\begin{equation}\label{TP1}
|\langle\Psi_m,\Psi_n(t)\rangle| = |\int\Psi_m(x+x_{nh}(t))\Psi_n(x)e^{-ix\dot{x}_{nh}(t)}dx|.
\end{equation}
To explicitly compute the scalar products we need to recall the following facts:
The eigenstate $\Psi_n(x)$ is expressed in terms of the Hermite polynomials as
\begin{equation}\label{F1}
\Psi_n(x) = \frac{1}{\sqrt{2^nn!}}\big(\frac{m\omega}{\pi}\big)^{1/4}H_n(\sqrt{m\omega} x)e^{-m\omega x^2/2}. 
\end{equation}
The constants are such that $\langle\Psi_n,\Psi_m\rangle=\delta_{n,m}.$ The Hermite polynomials are obtained from their generating function as follows:
\begin{equation}\label{F2}
G(x,u) = e^{2xu - u^2} = \sum_{n=0}^\infty \frac{u^n}{n!}H_n(x).
\end{equation}
To complete the list, we need to keep in mind that for any complex $z$ we have:
\begin{equation}\label{F3}
\int_{\bbR}e^{zx}e^{-x^2}dx = \sqrt{\pi}e^{\frac{z^2}{4}}.
\end{equation}.
If we leave aside the factor $C(m,n)=\Big(2^{(m+n)}m!n!\pi\Big)^{-1/2},$ and make the change of variables $\sqrt{m\omega}x\to x,$ to calculate the transition  probabilities we need to compute
\begin{equation}\label{TP2}
\int_{\bbR} H_m(x+a)H_n(x)e^{-ixb}e^{-(x+a)^2/2}e^{-x^2/2}dx
\end{equation}
\noindent where we put $a=\sqrt{m\omega}x_{nh}(t)$ and $b=\dot{x}_{nh}(t)/\sqrt{m\omega}.$ Instead of evaluation this integral for each $m,n$ we make use of (\ref{F2}) and evaluate instead
\begin{equation}\label{TP3}
\int_{\bbR}G(x+a,v)G(x,u)e^{-ixb}e^{-(x+a)^2/2}e^{-x^2/2}dx.
\end{equation}
To obtain the desired integral, we compute $\partial^{m+n}/\partial v^m\partial u^n$ at $v=u=0.$ The integrand in the last expression is an exponential, which after collecting powers of $x$ looks like
$$2av - a^2/2 - u^2 - v^2 + [2(u + v) - (ib + a)]x - x^2.$$
Now, make use of (\ref{F3}) to obtain
$$\int_{\bbR}G(x+a,v)G(x,u)e^{-ixb}e^{-x^2}dx = \sqrt{\pi}e^{2av-a^2/2-u^2-v^2}e^{[2(u+v)-(ib+a)]^2/4}.$$
At this point it is important to note that the square powers of $u$ and $v$ cancel out after the expansion and we are left with
\begin{equation}\label{TP4}
\begin{aligned}
\int_{\bbR}G(x+a,v)G(x,u)e^{-ixb}e^{-x^2}dx = \sqrt{\pi}\exp\big(2uv+2av -(u+v)(a+ib)-\frac{(a-ib)^2}{4}\big)\\
= \sqrt{\pi}\exp\big(2uv + v(a-ib) - u(a+ib) -\frac{(a-ib)^2}{4}\big).
\end{aligned}
\end{equation}
We can now replace back $a$ and $b$ with $\sqrt{\omega}x_{nh}(t)$ and $b=\dot{x}_{nh}(t)/\sqrt{\omega}.$ As a simple example we can compute the probability of finding the system in its ground state at time $t>0$ is given by 
\begin{equation}\label{TP5}
|\langle\Psi_0,\Psi_0(t)\rangle| = \exp\big(-\frac{m\omega x^2_{nh}(t)}{4} +\frac{x_{nh}(t)\dot{x}_{nh}(t}{2}-\frac{\dot{x}^2_{nh}(t)}{4m\omega}\big).
\end{equation}
The term in the exponent can be written as  $\big(\omega L_{nh}-\mathcal{E}_{nh}\big)/2\omega,$ where $L_{nh}$ in the angular momentum of the non-homogeneous solution and $\mathcal{E}_{nh}$ its total energy.
 
We direct the interested reader to \cite{BDQ} for a different approach to computing integrals of products of Hermite polynomials and exponentials. They use a somewhat different scaling for  the Hermite polynomials. 

\section{Closing comments}
We mentioned at the beginning that the time dependence of the forcing term is arbitrary, subject to the requirement that $\int_0^t|k(s)|ds$ be finite for all $t>0.$ But actually, at the expense of complicating the mathematical apparatus, we might consider impulsive forces (random or not), or white noise.

To finish and to repeat ourselves once more, sometimes a perturbed system can be rendered unperturbed by a change of coordinates. It is therefore convenient to consider a framework in which the equations of motion in both coordinate systems are the same. This is the role of canonical transformations in the Hamiltonian description of particle dynamics. The bonus is that in some cases the canonical transformation can be explicitly implemented as a unitary transformation between the quantized descriptions of the perturbed and unperturbed systems.

{\bf Acknowledgments} I want to thank the reviewers and the editors for their thorough and careful reading of the manuscript and for their comments and suggestions for improving it.

\end{document}